%% file: paper.tex
\documentclass{article}

\usepackage[final]{DLI_2023}

\usepackage[utf8]{inputenc} 
\usepackage[T1]{fontenc}    
\usepackage{hyperref}       
\usepackage{url}            
\usepackage{booktabs}       
\usepackage{amsfonts}       
\usepackage{nicefrac}       
\usepackage{microtype}      
\usepackage{xcolor}         
\usepackage{graphicx}

\title{Towards an AI to Win Ghana’s National Science and Maths Quiz}

\author{George Boateng\\
ETH Zurich\\
NSMQ AI, Kwame AI Inc.\\
\texttt{jojo@kwame.ai} \\
\And
Jonathan Abrefah Mensah \\
NSMQ AI, Kwame AI Inc. \\
\texttt{kwakuabrefahbusia@gmail.com} \\
\And
Kevin Takyi Yeboah \\
NSMQ AI, Kwame AI Inc. \\
\texttt{quakukevin@gmail.com} \\
\And
William Edor \\
NSMQ AI, Kwame AI Inc. \\
\texttt{edorwill@gmail.com} \\
\And
Andrew Kojo Mensah-Onumah \\
NSMQ AI, Kwame AI Inc. \\
\texttt{kojoakmo@gmail.com} \\
\And
Naafi Dasana Ibrahim \\
NSMQ AI, Kwame AI Inc. \\
\texttt{ibrahimnaafi@gmail.com} \\
\And
Nana Sam Yeboah \\
NSMQ AI, Kwame AI Inc. \\
\texttt{nanayeb34@gmail.com} \\
}

\begin{document}

\maketitle

\input{content.tex}

\clearpage
\bibliography{refs}
\bibliographystyle{paper}

\appendix
\clearpage
\input{appendix}

\end{document}

%% file: content.tex
\begin{abstract}
Can an AI win Ghana’s National Science and Maths Quiz (NSMQ)? That is the question we seek to answer in the NSMQ AI project, an open-source project that is building AI to compete live in the NSMQ and win. The NSMQ is an annual live science and mathematics competition for senior secondary school students in Ghana in which 3 teams of 2 students compete by answering questions across biology, chemistry, physics, and math in 5 rounds over 5 progressive stages until a winning team is crowned for that year. The NSMQ is an exciting live quiz competition with interesting technical challenges across speech-to-text, text-to-speech, question-answering, and human-computer interaction. In this ongoing work that began in January 2023, we give an overview of the project, describe each of the teams, progress made thus far, and the next steps toward our planned launch and debut of the AI in October for NSMQ 2023. An AI that conquers this grand challenge can have real-world impact on education such as enabling millions of students across Africa to have one-on-one learning support from this AI.
\end{abstract}

\section{Introduction}
According to UNESCO, 65\% of primary school teachers in Sub-Saharan Africa had the required minimum qualifications \cite{UNESCO2021}. Furthermore, the average student-teacher ratio at the primary education level in Sub-Saharan Africa in 2019 was 38:1 which is higher compared to 13.5:1  in Europe \cite{Eurostat2021}. Consequently, 15 million teachers are needed by 2030 across sub-Saharan Africa to reach education goals \cite{UNESCO2021b}, which is an expensive goal to accomplish. The lack of enough qualified teachers in Africa hampers adequate learning support for students. An Artificial Intelligence (AI) teaching assistant for teachers could potentially augment the efforts of the limited number of teachers in providing learning support such as one-on-one tutoring and question answering to students. Yet, there is currently no robust benchmark for real-world settings and the African context to evaluate such an AI.

Motivated by this need, in our position paper  — “\textit{Can an AI Win Ghana's National Science and Maths Quiz? An AI Grand Challenge for Education}” \cite{boateng2023} that was presented at the Practical Machine Learning (ML) for Developing Countries workshop at ICLR 2023, we proposed a grand challenge in education — \textbf{NSMQ AI Grand Challenge} — “\textit{Build an AI to compete in Ghana’s National Science and Maths Quiz Ghana (NSMQ) competition and win — performing better than the best contestants in all rounds and stages of the competition}”. We laid out the motivation for the work, key technical challenges, and future plans. That vision led to the creation of the NSMQ AI project, an open-source \footnote{NSMQ AI Open-Source Project: \url{https://github.com/nsmq-ai/nsmqai}} project to conquer this grand challenge. In this paper, we give an overview of the project, describe each of the teams, progress made thus far, and next steps toward our planned launch and debut of the AI in October for NSMQ 2023.

\section{Overview of Project}
The goal of the NSMQ AI project is to build an AI to win Ghana’s NSMQ. The NSMQ is an annual live science and mathematics competition for senior secondary school students in Ghana in which 3 teams of 2 students compete by answering questions across biology, chemistry, physics, and math in 5 rounds over 5 progressive stages until a winning team is crowned for that year.  The competition ran from 1993 to 2022 except for 2010 and 2011. The NSMQ is an exciting live quiz competition with interesting technical challenges across speech-to-text, text-to-speech, question-answering, and human-computer interaction. An AI that conquers this grand challenge can have real-world impact on education such as enabling millions of students across Africa to have one-on-one learning support from this AI.

Our goal is to have our AI compete in Round 5, the Riddles section of the 2023 competition happening in Fall 2023. This round, the final one, is arguably the most exciting round as the winner of the competition is generally determined by the performance in the round. In the Riddles round of the NSMQ quiz, students answer riddles across Biology, Chemistry, Physics, and Mathematics. Three (3) or more clues are read to the teams that compete against each other to be first to provide an answer (which is usually a word or a phrase) by ringing their bell. The clues start vague and get more specific. To make it more exciting and encourage educated risk-taking, answering on the 1st clue fetches 5 points, on the 2nd clue — 4 points, and on the 3rd or any clue thereafter, 3 points. There are 4 riddles in all for each contest with each riddle focusing on one of the 4 subjects. Speed and accuracy are key to winning the Riddles round. An example riddle with clues and the answer is as follows. \textbf{Question:} (1) I am a property of a periodic propagating disturbance. (2) Therefore, I am a property of a wave. (3) I describe a relationship that can exist between particle displacement and wave propagation direction in a mechanical wave. (4) I am only applicable to waves for which displacement is perpendicular to the direction of wave propagation. (5) I am that property of an electromagnetic wave which is demonstrated using a polaroid film. Who am I? \textbf{Answer:} Polarization.

The first version of the NSMQ AI \footnote{NSMQ AI System Demo: \url{https://youtu.be/rMqT4-kQVWM}} was demoed at AfricAIED 2023 \footnote{AfricAIED Workshop: \url{https://www.africaied.org}} — 1st Workshop on AI in Education, a satellite workshop of the International Conference on AI in Education. NSMQ AI is currently a web app that automatically transcribes speech with a Ghanaian accent from a video of a riddle, generates an answer to the scientific riddle and says that answer with a Ghanaian accent.

\section{NSMQ AI Teams}
The project has 6 teams that work together collaboratively towards accomplishing the goal: Data Curation, Data Preprocessing, Web App, Speech-to-Text, Question Answering, and Text-to-Speech.

\subsection{Data Curation}
The data curation team employs organizational skills in gathering and maintaining accurate, complete, and consistent datasets for the project. The data includes information on the contest date, names of the competing schools, the marks they acquired at the end of each contest, video clips of the contests, and video clips of riddles. For the riddles part, it includes information about which school answered the riddles and at which clue. All these are saved in a dedicated Google Sheet. Out of 11 years (2012-2022), we have substantial workable data for 5 years (2020-2018). Out of 71 contests for 2022, 68 of them have complete video links with 272 riddles. The years 2021, 2019, and 2018 have 40 contests each with 33, 39, and 34 complete video links and 148, 156, and 139 riddles respectively. The year 2020 is the only year filled with all its data information; 40 contests with 40 complete video links and 160 riddles. Work is ongoing to annotate the timestamps of each of the clues for every riddle in each contest. One key challenge is finding data sources. At the end of each contest, the competing schools and their respective scores are shared via images or PDFs on social media and blogs. Sometimes finding these materials is difficult as they may be scattered on the web or may not have been posted by the media channel in charge. Video recording errors also make getting accurate data difficult. Due to power surges and internet connection problems,  videos may have glitches, or the motion and the audio may overlap.

\subsection{Data Preprocessing}
The Data Preprocessing Team focuses on the automatic cleaning, transforming, and preparation of datasets. Additionally, the team provides essential support to the Data Curation Team through automation scripts, enhancing the efficiency of their work. The team developed a script for NSMQ video download and upload, streamlining the curation of NSMQ videos. Moreover, the team developed a riddles-cropping script to accurately extract riddle sections from NSMQ videos, facilitating easy utilization for training by the ML teams. The team also developed a script to extract HTML from subject-related books, segmenting content into chapters, sections, paragraphs, and passages, thereby providing valuable knowledge sources to the Question Answering Team. The following books have been parsed: High School Physics, College Physics, College Biology, College Chemistry, and College Algebra. Key challenges include ensuring data accuracy requiring meticulous verification, prompting close collaboration with the Data Curation Team to enhance data quality and inter-team cohesion, and continual optimization of automation scripts enabling efficient handling of larger datasets, ensuring scalability for future needs. 

\subsection{Web App}
The goal of the web app team is to develop an app that integrates the various machine learning subsystems and demonstrates their individual and collaborative operation within the NSMQ AI. The team has taken a user-centered approach in development where the interactions of the user have driven the frontend design and underlying backend architecture. The web app consists of demo and live quiz mode and currently serves as the coordinating layer of the NSMQ AI. It takes in user interaction and sends out API calls to servers running the ML inference code (currently on Google Colab). For example, in the demo mode, the user is able to provide real-time audio input on the  Speech-to-Text page. The web app processes this audio and sends an API call to the Speech-to-Text API server. The server returns the real-time transcript which is displayed back in the web app to the user. Utilizing this API approach allows each team to develop independently and scale as needed without impacting other subsystems. The web app utilizes Python in both the frontend and backend stack. The main framework used for development is Streamlit due to its applicability to ML and Data science use cases, quick development cycles, python language support which is more familiar to current and prospective team members, and simple cloud deployment approach using the Streamlit Community Cloud.


\subsection{Speech-to-Text}
The Speech-To-Text (STT) system \footnote{STT System Demo: \url{https://youtu.be/fM826DcUEIk}} is being built to provide a robust, fast service for Ghanaian-accented English in mathematical and scientific contexts. For an AI to outperform human competitors in the NSMQ competition, it is essential that the audio input received is quickly transcribed so further processing can be performed to derive answers faster than humans can. The STT system seeks to transcribe speech and select questions to be answered in real-time. We use OpenAI’s Whisper model \cite{radford2022}  for speech transcription and word error rate (WER) and latency as our evaluation metrics. We have built a system that performs transcription of real-time speech and existing audio. We have also built a pipeline to measure the performance of selected pre-trained Whisper models when transcribing audio. Also, the training of models with dummy data has been trialed, providing insights into key considerations for future training with data from the NSMQ competition. The main challenges faced are with respect to finding affordable resource allocations for model training and deployment. See Appendix \ref{sec:stt_eval} for initial experiments and evaluation results.

\subsection{Question Answering}
The Question Answering (QA) team’s primary goal is to develop a QA system \footnote{QA System Demo: \url{https://youtu.be/sDFlPzm6I98}} that answers NSMQ riddles swiftly and accurately and surpasses students. This undertaking presents a set of unique challenges that must be addressed to build an effective and competitive system. One critical challenge – among others – is determining the optimal moment for the model to return an answer, or deciding instead to wait for more information from additional clues to increase the model’s chances of providing the correct answer. Our current pipeline partially addresses these problems using two approaches. The first involves extracting contexts relevant to the current set of clues from a semantic search engine (Kwame AI \cite{boateng2022} or our custom-built engine using Sentence-BERT \cite{reimers2020}  and the Simple English Wikipedia dataset). The second, and core QA system, is based on a generative question-answering approach. We experimented with the Falcon-7b-instruct model to generate an answer given a set of riddle clues. We made use of prompt engineering to guide the model toward the desired output format. See Appendix \ref{sec:qa_eval} for initial experiments and evaluation results.

\subsection{Text-to-Speech}
The primary goal of the Text-To-Speech (TTS) team is to develop a TTS system \footnote{TTS System Demo: \url{https://youtu.be/raTWOo6aqk4}} that can synthesize answers from a Question-Answering (QA) system into audible speech with a Ghanaian accent. The challenges faced during development include limited data coverage with a Ghanaian accent, lack of support for scientific and mathematical speech synthesis, reliance on supervised learning, open source limitations, and finding a suitable evaluation metric for accent similarity. We used VITS \cite{kim2021}, an end-to-end model which we fine-tuned using voice samples from two Ghanaian speakers resulting in an output speech similar to those speakers \footnote{TTS Voice Sample: \url{https://youtu.be/dwg7izBMFGA}}. We used the VITS model because it achieves Mean Opinion Score closer to ground truth than other publicly available two-stage models. See Appendix \ref{sec:tts_eval} for initial experiments and evaluation results.

\section{Next Steps}
Over the following months ahead of our planned launch and debut of the AI in October for NSMQ 2023, we plan to address some of the key challenges and tasks that are still outstanding. For data curation, we plan to have audio segments (along with transcripts) of each riddle for at least 5 years of the competition (for use by the STT team) and the corresponding performance of the best human contestants (for benchmarking by the QA team). For Data Preprocessing, we plan to refine the HTML extraction validation process for better dataset output and eventually have a comprehensive corpus of science textbooks parsed and available for use by the QA team. For Web App, we plan to include the live mode where we will show the live stream of the contest as our AI attempts to answer in the Riddles round, make the app accessible on the cloud which involves cloud deployment and hosting to meet the system’s compute needs, reducing latency of ML inference subsystems as that can be an overall system bottleneck and supporting demo and live quiz modes which involve sample and live stream media data as well as user input. For STT, we plan to automatically detect the start of the reading of the riddles and to improve the accuracy and latency of the Whisper model by fine-tuning the model on past NSMQ audios and transcripts. For QA, we plan to improve the accuracy and latency of the model and generate confidence scores as the clues are being read to incorporate the model’s level of certainty into the question-answering process and inform attempts to answer early on. For TTS, we plan to improve the performance of scientific and mathematical speech synthesis and reduce the latency in synthesizing the speech output. We welcome all individuals interested in contributing to this open-source project to express interest by completing the following form \footnote{NSMQ AI Interest Form: \url{https://forms.gle/GTSrQxRN7pGWkbYK8}}.

%% file: appendix.tex
\section{Speech-To-Text Evaluation}
\label{sec:stt_eval}
For model performance evaluation, we tested our pipeline with 3 audios from the NSMQ competition which consist of speech with Ghanaian accents, along with their corresponding transcripts. Each audio was approximately 15 seconds long. The results from our evaluation of the pre-trained models are summarized in Table \ref{tab:stt_results}. For the evaluation metrics word error rate (WER) and latency, less is better. While a larger sample size would have been ideal, the evaluation provided tangible insights into models that could be used for custom training. However, we would have to perform further evaluation on a larger sample size to confirm all hypotheses.

The evaluation results significantly impacted our model selection for proof-of-concept training and deployment. The WERs are quite high, making the case to fine-tune the model with Ghanaian accented speech. None of the models attained the best scores for both metrics, which warranted the need for a trade-off between the model metrics. An ideal model would have the lowest word error rate and latency values. We also considered the unique demands of each proof-of-concept task in selecting the models. We settled on the small and medium models for training and deployment respectively. It is important to emphasize that our model selection used an iterative process, hence, further testing will be conducted to validate or correct our current assumptions.

\begin{table}[ht]
\caption{Evaluation results from Pre-Trained Whisper models}
\label{tab:stt_results}
\centering
\begin{tabular}{|l|l|l|}
\hline
\textbf{Model size}        & \textbf{Mean WER/\%}    & \textbf{Mean Latency/s} \\ \hline
Whisper tiny      & 31.11          & 2.88                    \\ \hline
Whisper tiny.en   & 30.29          & 1.77                    \\ \hline
Whisper base      & \textbf{29.70} & 1.33                    \\ \hline
Whisper base.en   & 30.69          & 1.06                    \\ \hline
Whisper small     & \textbf{29.51} & 0.97                    \\ \hline
Whisper small.en  & 31.29          & \textbf{0.88}           \\ \hline
Whisper medium    & 30.92          & \textbf{0.92}           \\ \hline
Whisper medium.en & 31.61          & 0.94                    \\ \hline
Whisper large     & 31.11          & 1.06                    \\ \hline
\end{tabular}
\end{table}

\section{Question Answering Evaluation}
\label{sec:qa_eval}
We used a dataset comprising a total of 1144 riddle-answer(s) pairs. The dataset was divided into train, test, and dev splits of 60:20:20. The train split comprises 686 riddle-answer(s) pairs, while each of the test and dev splits contains 229 riddle-answer(s) pairs. Each split was represented as a CSV file with the core columns “Clue 1” to “Clue 9” for all the clues in per riddle, “Answer” for the ground truth answer, and “Answer 1” to “Answer 4” for alternative ground truth answers (if any). We included additional columns to track information like subject, contest number and year of the contest. We performed evaluations on our test set. For each riddle-answer pair in the test set, we applied the following preprocessing steps before passing the content onto our evaluation pipeline: we converted all clue texts to lowercase, removed punctuation, and fixed whitespace. We did the same for answers and additionally, we removed articles (“the”, “a” “an”) as these have no bearing on the correctness of the final result.
For each riddle, we concatenated all clues that comprised the riddle into one string. For DistilBERT \cite{sanh2019} and SciBERT \cite{beltagy2019} (Extractive QA Models), we passed the concatenated clues into the semantic search engine to retrieve relevant passages. We retrieved the top three passages with the highest similarity scores, along with their confidence values. We concatenated the retrieved passages and calculated the mean of their confidence as the overall confidence. We then passed the concatenated clues into the extractive QA models as “questions”, while the retrieved passages are provided as the “context.” For Falcon-7b-Instruct (generative model), we formatted the concatenated clues into a carefully crafted prompt template and then passed this prompt to the model. For both classes of QA models, the generated answer per riddle was compared to the original ground truth answer, and if applicable, any alternative ground truth answers to the riddle. We performed this comparison in our evaluation pipeline. The evaluation pipeline assessed our models’ performance based on two metrics: Exact Match (EM), which checks for an exact match with ground truth answer(s), and Fuzzy Match (FM), which assesses the overlap of the generated answer with the ground truth answer(s). We evaluated our initial pipeline and benchmarked with GPT 3.5 \cite{chatgpt} (Table \ref{tab:qa_results}). Our generative model performed better than the extractive models but worse than ChatGPT. Finetuning our model on our dataset could potentially improve the performance.
\begin{table}[ht]
\label{tab:qa_results}
\caption{Evaluation results for QA Models}
\centering
\begin{tabular}{|l|c|c|c|}
\hline
 \textbf{Model} &  \textbf{Exact Match} &  \textbf{Fuzzy Match} \\
\hline
Falcon-7b-Instruct & \textbf{23.14\%} & \textbf{35.81\%} \\ \hline
DistilBERT & 8.3\% & 14.85\% \\ \hline
SciBERT & 0.0\% & 6.11\% \\ \hline
GPT-3.5 & 64.19\% & 75.54\% \\
\hline
\end{tabular}
\label{tab:performance_metrics}
\end{table}

\section{Text-To-Speech Evaluation}
\label{sec:tts_eval}
We performed two evaluation experiments and measured the Mean Opinion Score (MOS),  Word Error Rate (WER), and latency (Table \ref{tab:tts_results}). The automatic MOS is an objective evaluation of how ‘natural’ the synthesized speech sounds with a range from 1 (bad) to 5 (excellent). The WER of the synthesized speech measures the intelligibility of the synthesized speech, whereas the latency measures the inference speed of the model. The training datasets featured audio samples from two Ghanaians (with the permission of the speakers), each with unique recordings from TEDx talks and various sources like Youtube, podcasts, and presentations. Evaluation 1 involved synthesizing 30 scientific and mathematical speeches from past NSMQ questions and Evaluation 2 involved synthesizing 30 conversational speech. We compared the two VITS models (Vits-Elsie and Vits-Jojo) and benchmarked against two-stage models like Tacotron2 and Glow TTS. The results show that the Vits-Elsie model achieves the best WER (between Vits-Elsie and Vits-Jojo). The models generally exhibit high MOS scores, indicating a human-like sound. However, they struggle with scientific and mathematical text synthesis. 

\begin{table}[ht]
\centering
\caption{Evaluation results for TTS system}
\label{tab:tts_results}
\resizebox{\textwidth}{!}{%
\begin{tabular}{|l|ccc|ccc|}
\hline
 & \multicolumn{3}{c|}{\textbf{Evaluation 1}} & \multicolumn{3}{c|}{\textbf{Evaluation 2}} \\ \hline
\textbf{Model} & \multicolumn{1}{c|}{\textbf{Mean latency / s}} & \multicolumn{1}{c|}{\textbf{Mean WER (\%)}} & \textbf{Mean MOS} & \multicolumn{1}{c|}{\textbf{Mean latency / s}} & \multicolumn{1}{c|}{\textbf{Mean WER (\%)}} & \textbf{Mean MOS} \\ \hline
Fast pitch/Hifiganv2 & \multicolumn{1}{c|}{0.46} & \multicolumn{1}{c|}{\textbf{3.60}} & 2.98 & \multicolumn{1}{c|}{0.45} & \multicolumn{1}{c|}{\textbf{3.69}} & 2.98 \\ \hline
Tacotron2/Multiband Melgan & \multicolumn{1}{c|}{0.45} & \multicolumn{1}{c|}{4.14} & 2.99 & \multicolumn{1}{c|}{0.54} & \multicolumn{1}{c|}{4.14} & 2.99 \\ \hline
Glow TTs/Multiban Melgan & \multicolumn{1}{c|}{\textbf{0.26}} & \multicolumn{1}{c|}{6.32} & 2.99 & \multicolumn{1}{c|}{\textbf{0.26}} & \multicolumn{1}{c|}{6.32} & 2.99 \\ \hline
Vits\_Elsie & \multicolumn{1}{c|}{3.88} & \multicolumn{1}{c|}{35.76} & 3.38 & \multicolumn{1}{c|}{4.03} & \multicolumn{1}{c|}{5.40} & 3.20 \\ \hline
Vits\_Jojo & \multicolumn{1}{c|}{5.15} & \multicolumn{1}{c|}{88.47} & \textbf{3.39} & \multicolumn{1}{c|}{5.26} & \multicolumn{1}{c|}{63.07} & \textbf{3.198} \\ \hline
\end{tabular}
}
\end{table}